\vsize=19 cm
\hsize=12.5 cm
\voffset=1.9 cm
\hoffset=2 cm
\font\tinyrm=cmr8
\font\tinyit=cmti8
\font\tinybf=cmbx8
\font\tinysym=cmsy8
\rightline{Causality and Locality in Modern Physics, 171--178}
\rightline{G. Hunter, S. Jeffers, J.-P. Vigier  (eds.)}
\rightline{$\copyright$ 1998 {\it Kluwer Academic Publishers}}
\bigskip
\bigskip
\bigskip
\noindent {\bf THE ZERO-POINT FIELD AND INERTIA}
\bigskip
\bigskip
\parindent 2cm B. HAISCH\par {\it Solar and Astrophysics Laboratory, Lockheed
Martin}\par {\it 3251 Hanover St., Palo Alto, CA 94304}\par
\bigskip A. RUEDA\par {\it Dept. of Electrical Engineering \& Dept. of Physics}\par
{\it  California State Univ., Long Beach, CA 90840}

\parindent 0.25truein
\bigskip
\bigskip
\bigskip\noindent {\bf 1. Introduction}

\bigskip\noindent Is the vacuum electromagnetic zero-point field (ZPF) real? In a
nice collection of examples, Milonni (1988) has shown that the interpretations of
vacuum field fluctuations vs. radiation reaction are merely like two sides of the
same quantum mechanical coin (cf. Senitzky 1973). Physical phenomena such as
spontaneous emission, the Lamb shift and the Casimir force can be analyzed either way
with the same result. The Casimir force is of particular interest (Milonni 1982). The
recent measurements by Lamoreaux (1997) show agreement with the semiclassical theory
of Casimir based on a real ZPF to within 5\% over the measured range. Of course this
effect is derived in standard QED calculations via subtraction of two formally
infinite integrals over electromagnetic field modes. Another approach yielding
identical results simply treats the quantum vacuum as consisting of (virtual) photons
carrying linear momentum. Reflections off the conducting plates inside  and outside
the cavity are in balance for wavelengths shorter than the plate separation, but for
longer wavelengths modes are excluded within the cavity. This imbalance results in
a net ``zero-point radiation pressure'' pushing the plates together which is exactly
the Casimir force (Milonni, Cook \& Goggin 1988; Milonni 1994). {\it One may argue
over the correct theoretical perspective, but the new measurements leave no doubt
that the predicted macroscopic forces are quite real.}

A similar treatment of the quantum vacuum can be applied to scattering of
momentum-carrying zero-point photons by quarks and electrons in matter. As with the
Casimir cavity, such scattering is almost entirely a detailed balance process.
However it can be shown that, owing to acceleration effects within the class of those
first studied by Davies (1975) and Unruh (1976), an acceleration-dependent imbalance
results in a net reaction force (Rueda \& Haisch 1997, 1998). Thus as with the Casimir
force, a zero-point field quantum vacuum effect is proposed to give rise to a
macroscopic phenomenon: in this case, the inertia of matter.

\vfill\eject
\bigskip\noindent {\bf 2. The Zero-Point Field in Quantum Physics}

\medskip\noindent The Hamiltonian of a one-dimensional harmonic oscillator of unit
mass may be written (cf. Loudon 1983, chap. 4)

$${\hat H} = {1 \over 2} ({\hat p}^2 + \omega^2 {\hat q}^2 ) ,
\eqno(1)$$

\smallskip\noindent where $\hat p$ is the momentum operator and $\hat q$ the position
operator. From these the destruction (or lowering) and creation (or raising) operators
are formed:

$${\hat a} = (2 \hbar \omega)^{-1/2} (\omega {\hat q} + i {\hat p}) ,
\eqno(2a)$$

$${\hat a}^{\dag} = (2 \hbar \omega)^{-1/2} (\omega {\hat q} - i {\hat p}) .
\eqno(2b)$$

\smallskip\noindent The application of these operators to states of a quantum
oscillator results in lowering or raising of the state:

$${\hat a}              | n \rangle = n^{1/2} | n-1 \rangle , 
\eqno(3a)$$

$${\hat a}^{\dag} | n \rangle = (n+1)^{1/2}  | n+1 \rangle . 
\eqno(3b)$$

\smallskip\noindent Since the lowering operator produces zero when acting upon the 
ground state,

$${\hat a} | 0 \rangle = 0 , \eqno(4)$$

\smallskip\noindent the ground state energy of the quantum oscillator,
$| 0 \rangle$, must be greater than zero,

$${\hat H} | 0 \rangle = E_0 | 0 \rangle = {1 \over 2} \hbar \omega | 0 \rangle ,
\eqno(5)$$

\smallskip\noindent and thus for excited states

$$E_n = \left( n + {1 \over 2} \right) \hbar \omega . \eqno(6)$$

The electromagnetic field is quantized by associating a quantum mechanical harmonic
oscillator with each {\bf k}-mode. Plane electromagnetic waves propagating in a
direction {\bf k} may be written in terms of a vector potential
$\bf A_k$ as (ignoring polarization for simplicity)

$${\bf E_k} = i\omega_{\bf k} \{ {\bf A_k} {\rm exp} (-i\omega_{\bf k} t +i {\bf k
\cdot r}) - {\bf A_k^*} {\rm exp} (i\omega_{\bf k} t - i {\bf k \cdot r}) \} ,
\eqno(7a) $$
 
$${\bf B_k} = i{\bf k \times}   \{ {\bf A_k} {\rm exp} (-i\omega_{\bf k} t +i {\bf k
\cdot r})-{\bf A_k^* } {\rm exp} (i\omega_{\bf k} t -i {\bf k \cdot r}) \} .
\eqno(7b) $$

\smallskip\noindent Using generalized mode coordinates analogous to momentum ($P_{\bf
k}$) and position ($Q_{\bf k}$) in the manner of (2ab) above one can write $\bf A_k$
and $\bf A_k^*$ as

$${\bf A_k} = (4 \epsilon_0 V \omega_{\bf k}^2)^{-{1 \over 2}}
 (\omega_{\bf k} Q_{\bf k} + i P_{\bf k}) {\bf \varepsilon_k} , \eqno(8a)$$

$${\bf A_k^*} = (4 \epsilon_0 V \omega_{\bf k}^2)^{-{1 \over 2}}
 (\omega_{\bf k} Q_{\bf k} - i P_{\bf k}) {\bf \varepsilon_k} . \eqno(8b)$$

\smallskip\noindent In terms of these variables, the single-mode energy is

$$< E_{\bf k} > = {1 \over 2} (P_{\bf k}^2 + \omega_{\bf k}^2 Q_{\bf k}^2) .
\eqno(9)$$

\smallskip\noindent Equation (8) is analogous to (2), as is Equation (9) with  (1).
Just as mechanical quantization is done by replacing $\bf x$ and
$\bf p$ by quantum operators $\hat {\bf x}$ and $\hat {\bf p}$,  so is the
quantization of the electromagnetic field accomplished by replacing $\bf A$ with the
quantum operator $\hat {\bf A}$, which in turn converts $\bf E$ into the operator
$\hat {\bf E}$, and $\bf B$ into $\hat {\bf B}$.  In this way, the electromagnetic
field is quantized by associating each {\bf k}-mode (frequency, direction and
polarization) with a quantum-mechanical harmonic oscillator. The ground-state of the
quantized field has the energy

$$<E_{{\bf k},0}>  = {1 \over 2} (P_{{\bf k},0}^2 + \omega_{\bf k}^2 Q_{{\bf
k},0})^2  = {1 \over 2}  \hbar \omega_{\bf k} . \eqno(10)$$

\bigskip\noindent {\bf 3. The Zero-Point Field in Stochastic Electrodynamics}

\medskip\noindent Stochastic Electrodynamics (SED; see de la Pe\~na \& Cetto 1996;
Milonni 1994) treats the ZPF via a plane electromagnetic wave modes expansion
representation whose amplitudes are {\it exactly} such as to result in a
phase-averaged energy of
$\hbar
\omega /2$ in each mode ({\bf k},$\sigma$), where $\sigma$ represents polarization
(cf. Boyer 1975):

$${\bf E}^{ZP}({\bf r}, t) = {\rm Re} \sum_{\sigma=1}^2 \int d^3 k {\hat
\varepsilon}_{{\bf k},\sigma} \left[ {\hbar \omega_{\bf k} \over 8 \pi^3 \epsilon_0}
\right]^{1 \over 2} {\rm exp}(i{\bf k \cdot r}-i\omega_{\bf k} t +i\theta_{{\bf
k},\sigma}) , \eqno(11a)$$

$${\bf B}^{ZP}({\bf r}, t) = {\rm Re} \sum_{\sigma=1}^2 \int d^3 k ({\hat k} \times
{\hat
\varepsilon}_{{\bf k},\sigma}) \left[ {\hbar \omega_{\bf k} \over 8 \pi^3 \epsilon_0}
\right]^{1 \over 2} {\rm exp}(i{\bf k \cdot r}-i\omega_{\bf k} t +i\theta_{{\bf
k},\sigma}) . \eqno(11b)$$

\smallskip\noindent This kind of representation was used by Planck (1914) and
Einstein and co-workers (Bergia 1979). The stochasticity is entirely in the phase,
$\theta_{{\bf k},\sigma}$, of each wave. (As discussed in $\S$ 7 this is not entirely
correct.)

The spectral energy density of the classical ZPF is obtained from the number of modes
per unit volume, $8\pi \nu^2/c^3$ (Loudon 1983, Eq. 1.10), times the energy per mode,
$h\nu/2$. The Planck spectrum plus ZPF radiation is thus:

$$\rho(\nu,T)d\nu={8\pi\nu^2 \over c^3} \left( {h\nu \over e^{h\nu /kT} -1} +{h\nu
\over 2}\right) d\nu .
\eqno(12)
$$

\bigskip\noindent {\bf 4. The Davies-Unruh Effect}

\medskip\noindent Motivated by Hawking's evaporating black hole concept, Davies
(1975) and Unruh (1976)  determined that a Planck-like component of the
background scalar field will arise as seen from a
uniformly-accelerated point with constant proper acceleration {\bf a} (where
$|{\bf a}|=a$) having an effective temperature,

$$T_a = {\hbar a \over 2 \pi c k} .  \eqno(13)$$

\smallskip\noindent  (For the classical Bohr electron, $v^2/r \approx 10^{25}$
cm/s$^2$, $T_a \approx 367$ K.) This effect is derivable from quantum field theory
(Davies 1975, Unruh 1976). It was also derived in SED for the classical ZPF by Boyer
(1980) who obtained for the spectrum a quasi-Planckian form (in the absence of
external radiation):

$$\rho(\nu,T_a)d\nu = {8\pi\nu^2 \over c^3}
\left[ 1 + \left( {a \over 2 \pi c \nu} \right) ^2 \right]
\left[ {h\nu \over 2} + {h\nu \over e^{h\nu/kT_a}-1} \right] d\nu  . \eqno(14)$$

\bigskip\noindent {\bf 5. Newtonian Inertia from ZPF Electrodynamics}

\medskip\noindent While these additional acceleration-dependent terms in Eq. (14) do
not show any spatial asymmetry in the expression for the ZPF spectral energy density,
certain asymmetries do appear when the electromagnetic field interactions with
charged particles are analyzed. 
Haisch, Rueda \& Puthoff (HRP; 1994) made use of this to propose a connection between
the ZPF and inertia of matter. Assume that there are interactions between a real ZPF,
represented as above, and matter at the fundamental particle level, treated as a
collection of electrons and quarks, both of which are simply thought of as oscillating
point charges: partons in the terminology of Feynmann. If the ZPF-parton interactions
take place at high frequencies, then one need not worry about how the three quarks in
a proton or a neutron are bound together. Each will interact independently with the
ZPF, even though the three-quark ensemble is constrained to macroscopically move
together.

The method of Einstein and Hopf (1910) was followed: it breaks the analysis of the
dynamics of the uniformly-accelerated parton into two steps. First we assume that the
electric component of the ZPF,
${\bf E}^{zp}$, drives the parton to harmonic oscillation, i.e. creates a Planck
oscillator. For simplicity we restrict these oscillations to a $yz$-plane
characterized by the velocity vector
${\bf v}_{osc}$ and we force the oscillating parton to accelerate, via an external
agent, in the
$x$-direction with constant acceleration ${\bf a}$ (perpendicular to ${\bf v}_{osc}$).
The acceleration will introduce asymmetries in the ZPF radiation field perceived by
the oscillating parton. Second, we then ask what the effect is of the magnetic
ZPF-parton interactions, specifically, what is the resulting Lorentz force:
$<{\bf v}_{osc}\times {\bf B}^{zp}>$? The result was the discovery of a reaction
force of the form

$${\bf F}_r = \left< {\bf v}_{osc}\times {\bf B}^{zp} \right> = -\left[ {\Gamma_Z
\hbar
\omega_c^2
\over 2\pi c^2} \right] {\bf a}.
\eqno(15)$$

\noindent The quantity in brackets on the right hand side we interpreted as the
inertial mass,

$$m_i = \left[ {\Gamma_Z \hbar \omega_c^2 \over 2\pi c^2} \right]
\eqno(16)$$

\noindent where $\Gamma_Z$ is the classical radiation damping constant of Abraham and
Lorentz, but now referring to the Zitterbewegung oscillations
\footnote{$^c$} {\tinyrm As discussed in chapter 17 of Jackson (1975) {\tinyit
Classical Electrodynamics}, one can obtain a characteristic radiation damping time
for an electron having the value
$\Gamma_e = 6.26 \times 10^{-24}$. This is not the proper $\Gamma_Z$ for
Zitterbewegung.} and $\omega_c$ was taken to be an effective cut-off frequency of
either the ZPF spectrum itself (perhaps at the Planck frequency) or of the
particle-field interaction owing to a minimum (Planck) particle size (Rueda 1981).
Newton's Third Law tells us that a (motive) force, ${\bf F}$ will generate an equal
and opposite (reaction) force, ${\bf F}_r$, and from Equation (15)

$${\bf F} = -{\bf F}_r = m_i {\bf a}.
\eqno(17)$$

\noindent Newton's third law is fundamental, whereas
Newton's second law, ${\bf F}=m{\bf a}$, appears to be derivable from the
third law together with the laws of electrodynamics.

\bigskip\noindent {\bf 6. The Relativistic Formulation of ZPF-based Inertia}

\medskip\noindent The oversimplification of an idealized oscillator
interacting with the ZPF as well as the mathematical complexity of the HRP analysis
are understandable sources of skepticism, as is the limitation to Newtonian
mechanics. A relativistic form of the equation of motion having standard covariant
properties has been obtained (Rueda \& Haisch 1998), which is independent of any
particle model, relying solely on the standard Lorentz-transformation properties of
the electromagnetic fields.

Newton's third law states that if an agent applies a force to a point on an object, at
that point there arises an equal and opposite force back upon the agent. Were this
not the case, the agent would not experience the process of exerting a force and we
would have no basis for mechanics. The mechanical law of equal and opposite contact
forces is thus fundamental both conceptually and perceptually, but it is legitimate
to seek further underlying connections. In the case of a stationary object (fixed to
the earth, say), the equal and opposite force can be said to arise in interatomic
forces in the neighborhood of the point of contact which act to resist compression.
This can be traced more deeply still to electromagnetic interactions involving
orbital electrons of adjacent atoms or molecules, etc.

A similar experience of equal and opposite forces arises in the process of
accelerating (pushing on) an object that is free to move.  It is an experimental fact
that to accelerate an object a force must be applied by an agent and that the agent
will thus experience an equal and opposite reaction force so long as the acceleration
continues.  It appears that this equal and opposite reaction force also has a deeper
physical cause, which turns out to also be electromagnetic and is specifically due to
the scattering of ZPF radiation. Rueda \& Haisch (1998) demonstrate that from the
point of view of the pushing agent there exists a net flux (Poynting vector) of ZPF
radiation transiting the accelerating object in a direction opposite to
the acceleration. The scattering opacity of the object to the transiting flux
creates the back reaction force called inertia.

The new approach is less complex and model-dependent than the HRP analysis in that it
assumes simply that the fundamental particles in any material object interact with
the ZPF in some way that is analogous to ordinary scattering of radiation. It is well
known that treating the ZPF-particle interaction as dipole scattering is a successful
representation in that the dipole-scattered field exactly reproduces the original
unscattered field radiation pattern, i.e. results in detailed balance. It is thus
likely that dipole scattering is a correct way to describe the ZPF-particle
interaction, but in fact for our analysis we simply need to assume that there is some
dimensionless efficiency factor,
$\eta(\omega)$, that describes whatever the process is (be it dipole scattering or
not). We suspect that $\eta(\omega)$
contains one or more resonances, but again this is not a necessary assumption.

The new approach relies on making standard transformations of the
${\bf E}^{zp}$ and ${\bf B}^{zp}$ from a stationary to an accelerated coordinate
system (cf. $\S$ 11.10 of Jackson, 1975). In a stationary or uniformly-moving frame
the ${\bf E}^{zp}$ and
${\bf B}^{zp}$ constitute an isotropic radiation pattern. In an accelerated frame the
radiation pattern acquires asymmetries. There is thus a non-zero Poynting vector in
any accelerated frame carrying a non-zero net flux of electromagnetic momentum. The
scattering of this momentum flux generates a reaction force, ${\bf F}_r$. Moreover
since any physical object will undergo a Lorentz contraction in the direction of
motion the reaction force,
${\bf F}_r$, can be shown to depend on $\gamma_{\tau}$, the Lorentz factor (which is a
function of proper time, $\tau$, since the object is accelerating). We find that

$$m_i=\left[ {V_0 \over c^2} \int \eta(\omega) {\hbar \omega^3 \over 2\pi^2c^3 }
d\omega
\right] .
\eqno(18)$$

\noindent We find the momentum of the object to be of the form

$${\bf p}=m_i \gamma_{\tau} {\bf v}_{\tau} .
\eqno(19)$$

\noindent Thus, we arrive at the relativistic equation of motion

$${\cal F}={d{\cal P} \over d\tau} = {d \over d\tau} (\gamma_{\tau} m_i c, \ {\bf p}
\ ) .
\eqno(20)$$

\indent The origin of inertia becomes remarkably intuitive. Any material object
resists acceleration because the acceleration produces a perceived flux of radiation
in the opposite direction that scatters within the object and thereby pushes against
the accelerating agent. Inertia is a kind of acceleration-dependent electromagnetic
drag force acting upon fundamental charges particles.

\bigskip\noindent {\bf  7. For the Future}

\medskip\noindent Clearly a quantum field theoretical derivation of the ZPF-inertia
connection is highly desireable. Another approach would be to demonstrate the exact
equivalence of SED and QED. However as shown convincingly by de la Pe\~na and Cetto
(1996), the present form of SED is not compatible with QED, but modified forms could
well be, such as their own proposed ``linear SED.'' Another step in the direction of
reconciling SED and QED is the proposed modification of SED by Ibison and Haisch
(1996), who showed that a modification of the standard ZPF representation (Eqs. 11ab)
can exactly reproduce the statistics of the electromagnetic vacuum of QED.

For an oscillator of amplitude $\pm A$ the classical probability of finding the
point-mass in the interval $dx$ is a smooth function with a minimum at the origin
(where the velocity is greatest) and a maximum at the endpoints of the oscillation.
Treated quantum mechanically, an oscillator has a very different behaviour, but in an
excited state approximates the classical probability distribution in the mean (see
Fig. 1 of Ibison \& Haisch). However the quantum $n=0$ ground-state --- the one of
direct relevance to the ZPF ---  is radically different from the classical one:  the
quantum-state probability maximum occurs where the classical state probability is at
a minimum (position zero) and vice versa at the endpoints; indeed the quantum
probability distribution is non-zero beyond
$\pm A$.  In both cases the average position, of course, remains zero.
The same disagreement characterizes the difference between the Boyer description of
the ZPF and the quantum ZPF.  It has been  shown by Ibison and Haisch (1996) that
this can be remedied by introduction of a  stochastic element into the amplitude of
each mode that precisely agrees with the quantum statistics. This gives us confidence
that the SED basis of the inertia (and gravitation) concepts is a valid one.

The two most frequently posed questions, indeed perhaps the most important ones, are
(1) whether the ZPF-inertia theory is subject to experimental validation, and (2)
what the implications might be for revolutionary new technologies. An independent
assessment of the case for experimental testing was carried out by Forward (1996) as
a USAF-sponsored study. No {\it direct} test could be identified as currently
feasible, but a constellation of related experiments were identified.

A NASA Breakthrough Propulsion Physics program is being initiated, and the
ZPF-inertia concept is high on the list of candidate ideas to explore (see Haisch \&
Rueda 1997a) along with the Sakharov-Puthoff concept of ZPF-gravitation linked to the
ideas herein by the principle of equivalence (Sakharov 1968, Puthoff 1989, see also
Haisch \& Rueda 1997b). We note that four decades elapsed before atomic energy became
a technology following Einstein's 1905 paper proposing (special) relativity. A
similar time-scale may apply here.

\bigskip\noindent {\bf  Acknowledgements}

\medskip\noindent We acknowledge support of NASA contract NASW-5050 for this work. BH
also thanks Prof. J. Tr\"umper and the Max-Planck-Institut where some of these ideas
originated during several extended stays as a Visiting Fellow. AR acknowledges
valuable discussions with Dr. D. C. Cole.

{

\bigskip
\parskip=0pt plus 2pt minus 1pt\leftskip=0.25in\parindent=-.25in

{\bf References}
{\tinyrm 
\medskip

Bergia, S. (1979), The light quantum and the wave-particle duality for radiation in
Einstein's research: 1904--1925, preprint, Inst. di Fisica, U. Bologna.

Boyer, T.H. (1975), Random electrodynamics: The theory of classical electrodynamics
with classical electromagnetic zero-point radiation, {\tinyit Phys. Rev. D}, {\tinybf
11}, 790--808.

Boyer, T.H. (1980), Thermal effects of acceleration through random classical
radiation, {\tinyit Phys. Rev. D}, {\tinybf 21}, 2137--2148.

Boyer, T.H. (1984), Thermal effects of acceleration for a classical dipole
oscillator in classical electromagnetic zero-point radiation, {\tinyit Phys. Rev.
D}, {\tinybf 29}, 1089--1095.

Davies, P.C.W. (1975), Scalar particle production in Schwarzschild and Rindler
metrics, {\tinyit J. Phys. A}, {\tinybf 8}, 609.

de la Pe\~na, L. \& Cetto, A. (1996), {\tinyit The Quantum Dice: An Introduction to
Stochastic Electrodynamics}, (Kluwer Acad. Publ., Dordrecht).

Einstein, A. \& Hopf, L. (1910), \"Uber einen Satz der
Wahrscheinlichkeitsrechnung und seine Anwendung in der Strahlungstheorie, {\tinyit
Annalen der Physik (Leipzig)}, {\tinybf 33}, 1096; Statistische Untersuchung
der Bewegung eines Resonators in einem Strahlungsfeld, {\tinybf 33}, 1105.

Forward, R.L. (1996), Mass Modification Experiment Definition Study, Phillips
Lab. Rept. 96-3004; also published in {\tinyit J. Sci. Expl.}, {\tinybf 10}, 325--354
(1996).

Haisch, B. \& Rueda, A. (1997a), The Zero-Point Field and the NASA Challenge to
Create the Space Drive, {\tinyit Proc. NASA Breakthrough Prop. Physics Workshop},
in press.

Haisch, B. \& Rueda, A. (1997b), Reply to Michel's `Comment on Zero-Point
Fluctuations and the Cosmological Constant', {\tinyit Ap. J.}, {\tinybf 488}, 563.

Haisch, B., Rueda, A. \& Puthoff, H.E. (1994), Inertia as a zero-point field
Lorentz Force, {\tinyit Phys. Rev. A}, {\tinybf 49}, 678.

Ibison, M. \& Haisch, B. (1996), Quantum and classical statistics of the
electromagnetic zero-point field, {\tinyit Phys. Rev. A}, {\tinybf 54}, 2737.

Jackson, J. D. (1975), {\tinyit Classical Electrodynamics}, 2nd ed., Wiley \& Sons.

Lamoreaux, S.K. (1997), Demonstration of the Casimir Force in the 0.6 to 6
${\tinysym \mu}$m Range, {\tinyit Phys. Rev. Letters} {\tinybf 78}, 5--8.

Loudon, R. (1983), {\tinyit The Quantum Theory of Light (2nd. ed).}, Oxford Univ.
Press, chap. 4.

Milonni, P.W. (1982), Casimir forces without the radiation field, {\tinyit Phys. Rev.
A} {\tinybf 25}, 1315--1327.

Milonni, P.W. (1988), Different Ways of Looking at the Electromagnetic Vacuum,
{\tinyit Physica Scripta} {\tinybf T21}, 102--109..

Milonni, P.W. (1994), {\tinyit The Quantum Vacuum}, Academic Pres, chap. 1.

Milonni, P.W., Cook, R.J. \& Goggin, M.E. (1988), Radiation pressure from the vacuum:
Physical interpretation of the Casimir force, {\tinyit Phys. Rev. A} {\tinybf 38},
1621--1623.

Planck, M. (1914), The theory of heat radiation (P. Blakansten \& Son, London)

Puthoff, H.E. (1989), Gravity as a zero-point fluctuation force, {\tinyit Phys.
Rev. A}, {\tinybf 39}, 2333.

Rueda, A. (1981), Behaviour of classical particles immersed in the classical
electromagnetic field, {\tinyit Phys. Rev. A}, {\tinybf 23}, 2020.

Rueda, A. \& Haisch, B. (1997), Electromagnetic vacuum and inertial mass, this
volume.

Rueda, A. \& Haisch, B. (1998), Inertial mass as reaction of the vacuum to
accelerated motion {\tinyit Phys. Lett. A}, {\tinybf 240}, 115.

Sakharov, A. (1968), Vacuum Quantum Fluctuations in Curved Space and the Theory of
Gravitation {\tinyit Soviet Physics - Doklady}, {\tinybf 12}, No. 11, 1040.

Senitzky, I.R. (1973), Radiation reaction and vacuum field effects in Heisenberg
picture QED, {\tinyit Phys. Rev. Lett.}, {\tinybf 31}, 955.

Unruh, W.G. (1976) Notes on black-hole evaporation, {\tinyit Phys. Rev. D}, {\tinybf
14}, 870.

}}

\bye